\documentstyle[12pt,epsfig]{article}
\begin{document}
\renewcommand{\theequation}{\thesection.\arabic{equation}}

%\mbox{ } \hfill version ( May 30th, 1995 )
\vspace*{2cm}
\begin{center}
{\LARGE\bf t-Channel Approach to Reggeon Interactions in QCD   }
\end{center}
\vspace{2cm}
\begin{center}
{\large\bf R. Kirschner \footnote{\it supported by Deutsche
Forschungsgemeinschaft}}

{\large\it
Institut f\"ur Theoretische Physik,  Universit\"at Leipzig,}
\\{\large\it D-04109 Leipzig, Germany }
\end{center}
\vspace*{5.0cm}
\begin{center}
{\bf Abstract }
\end{center}

Starting from the multi-Regge effective action for high-energy
scattering in QCD a $t$-channel approach can be developed which
is similar to the approach by White based on general Regge
arguments. The BFKL kernel of reggeized gluon interaction,
contributions to the $2 \rightarrow 4 $ reggeized gluon
vertex function and the one-loop correction to the BFKL
kernel are considered. The conditions are discussed under
which this approach can provide a simple estimante of the
next-to-leading logarithmic corrections to the BFKL
perturbative pomeron intercept.

\newpage
\section{Introduction}

Together with the progress in the experiments at HERA on deep-inelastic
scattering at small $x$ \cite{HERA} the ideas and the results about the
perturbative
Regge asymptotics became popular. Long-standing results about the
perturbative pomeron \cite{BFKL} \cite{ChLo} have found here a
successful phenomenological application. Quite some effort has been
applied in recent years to work out the corrections to the leading log
results and also the predictions about distributions in the hadronic
final state of small-$x$ deep-inelastic scattering.

We consider the part of the Regge region where the typical transverse
momenta are large compared to the hadronic scale. It is natural and
extremely useful to rewrite the leading perturbative contributions in
this region in terms of the exchange of reggeized gluons, which interact
among each other \cite{BFKL} \cite{ChLo} \cite{JB} \cite{KwP}
\cite{ARW} \cite{Levrev}.

A particular useful formulation is provided by the multi-Regge effective
action \cite{Lev90} \cite{KLS} \cite{Lev95}, which can be obtained
directly from the original QCD action by integration over modes
of the fields, corresponding to momenta outside the kinematical  range
of scattering and exchanged gluons and quarks. Here we shall not be
concerned with the case of quark exchange.

In this paper we discuss the relations of the multi-Regge effective
action to a reggeon approach \cite{AWCC1} \cite{AWCC2} which reproduces
the leading log reggeon interaction (BFKL kernel) and claims to predict
some part of the higher loop corrections. The approach is based on an
understanding about the connection between general Regge concepts to
the perturbative and non-perturbative properties of non-abelian gauge
theories developed by  White in \cite{ARW} (and references therein).
The
essential ingredients of the reconstruction procedure are signature-odd
reggeons, related to the exchanged gluons,  and their triple vertex,
characterized by the non-sense zero behaviour:
\begin{eqnarray}
 \Gamma_{1,2} (\kappa, \kappa_1 , \kappa_2 )
&= g ( \omega - \alpha^{\prime} \vert \kappa_1 \vert^2
 - \alpha^{\prime} \vert \kappa_2 \vert^2  )
\cr
&\sim (\vert \kappa \vert^2 - \vert \kappa_1 \vert^2  - \vert \kappa_2
\vert^2 )  \ \ {\rm for } \ \ \omega = \alpha^{\prime } \vert \kappa
\vert^2 , \cr
&\sim \vert \kappa \vert^2 \ \   {\rm for } \ \ \ \kappa_{1,2}
\rightarrow 0  .
\end{eqnarray}
$\alpha^{\prime } $ is the slope of the Regge trajectory, which is taken
to zero at the
end. $ \omega = \alpha^{\prime } \vert \kappa \vert^2 $ is the position
of the pole in the $\omega $-plane (here $\omega +1 $ is the angular
momentum
) of the incoming reggeon propagator. Remarkably, those reggeons
are colour singlets and the vertex has a trivial gauge-group structure.

Transitions of $m$ to $n$ reggeons in the $t$-channel are constructed
step by step starting with the vertex (1.1) and applying a condition
related to the Ward identities of the gauge theory:

$n \rightarrow m$ reggeon Green function vanish together with the
transverse momentum of any in- or outgoing reggeon. Higher vertices are
introduced additionally to (1.1) in order to obey this condition.

The property of vanishing at vanishing transverse momenta and its
relation to gauge invariance has been noticed also in \cite{Lev86}.

We point out that the triple vertex of exchanged gluons, emerging in the
effective action as the induced vertex from the integration over heavy
modes, has features close to the ones of (1.1). Unlike White's reggeon
vertex it  carries the gauge-group structure of the ordinary
triple-gluon vertex and depends on longitudinal momenta.
We explain how this vertex can be used to reproduce the leading log
results about the gluon reggeization and the (BFKL) interaction kernel
in a t-channel approach close to \cite{AWCC1}.
The leading contribution to the $2 \rightarrow 4$ QCD reggeon vertex
has been constructed by  Bartels \cite{JB93}.
There the transition kernels $K_{2 \rightarrow n}, (n = 2,3,4) $
obtained earlier \cite{JB} are the essential building blocks.
We obtain these transition kernel in two ways from the multi-Regge
effective action: From
the effective scattering and production vertices (s-channel approach)
and from the triple vertex of exchanged gluons (t-channel approach).

We discuss in detail the conditions on the gauge group state and the
longitudinal momenta of the $t$-channel intermediate state under which
the triple vertex of exchanged gluons can result in the ${\cal O} (g^4)
$ kernel proposed in \cite{AWCC1} as the next-to-leading log correction
to the QCD reggeon interaction (BFKL) kernel. From this it becomes clear
in which sense the results of \cite{AWCC1} \cite{AWCC2} can approximate
the full next-to-leading QCD calculations, some part of which has been
done already \cite{FL}, and whether the t-channel reggeon approach can
become a tool to estimate higher order terms. In particular we point out
a factor which is necessary to estimate the correction by using this
${\cal O}(g^4) $ kernel.

Finally we outline our method of calculating the eigenvalue of the
${\cal O}(g^4) $ kernel. The result is known and a method of calculation
has been described in a recent paper \cite{AWCC2}. That calculation is
based on advanced methods \cite{KT}. Our approach is just a sequence of
simple steps like the decomposition into partial fractions and
the representation of cut-off regularization by complex contour
integrals.

\section{The triple vertex of exchanged gluons }
\setcounter{equation}{0}

The leading contribution to scattering processes in the perturbative
Regge region which obey the multi-Regge kinematics in all sub-energy
channels is described by the following effective action (restricted to
gluons only):

\begin{eqnarray}
{\cal L} = {\cal L}_{kin} + {\cal L}_{s} + {\cal L}_{p} + {\cal L}_{t},
\cr
  {\cal L}_{kin} =
  - \frac{1}{2} (\partial^* \phi^a)
       \Box (\partial \phi^{a*})
  - 2 {\cal A}^a_+ \partial\partial^*
       {\cal A}^a_-  ,  \cr
{\cal L}_{s} =  - \frac{i g}{2}
          (\partial \phi^*  T^a \stackrel{\leftrightarrow }\partial_-
\partial^* \phi) {\cal A}^{a}_+
   +   \frac{i g}{2} (
     \partial^* \phi^*) T^a \stackrel{\leftrightarrow }\partial_+
 \partial \phi ) A^{a}_- ,
       \nonumber \\
{\cal L}_p = g  \phi^a(\partial {\cal A}_- T^a \partial^*
      {\cal A}_+ ) + g  \phi^{a *} (\partial^*
     {\cal A}_- T^a  \partial {\cal A}_+ ) ,   \nonumber  \\
    {\cal L}_{t} = \frac{ig}{2} \partial \partial^* {\cal A}^{a}_-
(\partial_+^{-1} {\cal A}_+ T^{a} {\cal A}_+ )
+ \frac{ig}{2}\partial \partial^* {\cal A}^{a}_+ (\partial_-^{-1} {\cal
A}_- T^{a} {\cal A}_- )
\end{eqnarray}
We use the notations as in \cite{KLS}. Four-vectors are decomposed into
the two-dimensional components parallel (longitudinal) and orthogonal
(transverse) to the plane determined by the (light-like) momentum
vectors $p_A , p_B $ of the incoming particles. The longitudinal
component is parametrized by the light-cone coordinates and the
transverse component is represented by the corresponding complex number.
\begin{eqnarray}
k_{\pm } = k_0 \pm k_3,  \kappa = k^1 + i k^2, \kappa^* = k^1 - i k^2,
\cr
x_{\pm} = x_0 \pm x_3 , x = x^1 + i x^2, x^{*} = x^1 - i x^2,
\cr
\partial_{\pm } = \frac{1}{2} (\partial_0 \pm \partial_3 ),
\partial = \frac{1}{2} (\partial_1 - i \partial_2 ),
\partial^* =  (\partial )^*.
\end{eqnarray}
Further we use the abbreviation
\begin{equation}
(A T^{a} B) = -i f^{abc} A^{b} B^{c} .
\end{equation}

The complex scalar field $\phi (x) $ represents the scattering gluons
with their two helicity states. Correspondingly the momentum modes are
restricted to the range close to mass shell.

\begin{equation}
  \phi :  \vert k_+ k_-  - \vert \kappa \vert^2 \vert  \ll q_0^2 .
\end{equation}

${\cal A}_{\pm} $ represent the exchanged gluons going from scattering
particles with large $k_{\mp} $ to scattering particles with large
$k_{\pm} $. Their momentum modes are restricted to the range

\begin{equation}
{\cal A}_{\pm }: \vert k_+ k_- \vert \ll \vert \kappa \vert^2  \sim
q_0^2.
\end{equation}
Their propagator depends only on transverse momenta, as follows formally
from (2.1), if the longitudinal momenta between which they are exchanged
are ordered as described. According to (2.5) we have to
adopt additionally that their
propagator vanishes, if this longitudinal momentum ordering is violated.

Both $\phi^{a} (x) $ and $ {\cal A}^{a}_{\pm} $ are directly related to
the original gauge potential $A^{a}_{\mu } $. The restrictions on the
momentum modes have been used extensively in deriving (2.1), in
particular the modes with
$ \vert k_+ k_- \vert \gg \vert \kappa \vert^2 $
have been integrated
out. The restrictions have to be imposed as additional conditions
when constructing the leading log amplitudes from (2.1). When one
relaxes
the condition (2.4) on $\phi $ one can disregard the triple interaction
of exchanged gluons ${\cal L}_{t} $ and the cut-off $q_0 $ does never
appear.

The vertices in ${\cal L}_p , {\cal L}_s $ and ${\cal L}_t $ are
represented by graphs as in Fig. 1. Here (unlike in \cite{KLS}) the
dotted lines represent the scattering gluons $\phi$ and the full lines
the exchanged gluons ${\cal A}_{\pm} $. The $t$-channel is drawn in the
horizontal direction. Scattering particles on the left (right) side
carry large $k_-  (k_+ ) $.

The exchanged gluons ${\cal A}_{\pm} $ have features reminicent to
reggeons. One such feature is their orientation $(\pm)$ related to the
longitudinal momentum ordering. The important point for this paper is
the observation that their triple vertices  ${\cal L }_{t} $ have a
transverse momentum dependence close to the non-sense zero structure
(1.1) of the triple odd-signature reggeon vertex \cite{ARW}
\cite{AWCC1}.
For the vertex ${\cal A}_- \rightarrow {\cal A}_+  {\cal A}_+ $, Fig.
1c, we have in the momentum representation
\begin{equation}
i g f^{abc} { \vert \kappa_1 + \kappa_2 \vert^2 \over k_{2 +} } .
\end{equation}
Notice that due to the longitudinal momentum ordering $k_{2 +} \approx -
k_{1 +} $  and within this accuracy the vertex is symmetric under
exchanging $ (b, k_2 ) \leftrightarrow (c, k_1 )$. The vertex changes an
odd-signature state (single exchanged gluon) into an even-signature
state.

 It is well known that the exchanged gluons reggeize. We repeat the
corresponding discussion here in order to emphasize the role of the
vertex ${\cal L}_t $
\footnote{The role of this vertex
in reggeization was clear to the authors of \cite{KLS}, but the papers
contain only short comments.}
in comparison with the role of the effective vertex
of gluon production ${\cal L}_p $.

Consider a state in the t-channel of one or two gluons corresponding to
the adjoint representation. The leading log contributions to the
amplitude with this exchange channel is determined by the graphs of Fig.
2
 iterated in the $t$-channel direction. Iterating the box graph Fig. 2a
only,  constructed with the effective production vertices, one obtains
the full contribution, if one removes the cut-off $q_0 $ in (2.2), $q_0
\rightarrow \infty $. More precisely, the sum of ladder graphs built
with the effective vertex and with reggeized gluons in the $t$-channel
results in  a single reggeized gluon exchange (bootstrap).

Consider
for arbitrary $q_0 $ the loop integral corresponding to Fig. 2b,
\begin{eqnarray}
{g^2 N \over 2 (2 \pi )^3 } \int {dk_+ dk_- \over k_+ k_- }
\int {d^2\kappa^{\prime } \vert \kappa \vert^4 \over
\vert \kappa^{\prime } \vert^2
\vert \kappa - \kappa^{\prime } \vert^2 } .
\end{eqnarray}
We specify below the regularization necessary to give the transverse
momentum integral a meaning. It is proportional to the deviation $ -
g^2 N \alpha (\kappa ) $ of the gluon trajectory from the angular
momentum
value $j = 1$.
\begin{equation}
\alpha (\kappa ) =
{1 \over 2 (2 \pi )^3 } \int {d^2\kappa^{\prime } \vert \kappa \vert^2
\over \vert \kappa^{\prime } \vert^2
\vert \kappa - \kappa^{\prime } \vert^2 } .
\end{equation}

To understand the longitudinal momentum integral
we restore the heavy-mode intermediate state \cite{KLS} from which the
triple vertex of exchanged gluons ${\cal L}_t $ emerged. The heavy line
in Fig. 3b
 represents the virtual gluon modes with
\begin{equation}
\vert k_+ k_- - \vert \kappa \vert^2 \vert  \gg q_0^2 .
\end{equation}
 This step is similar to the way one understands reggeon diagrams of the
form Fig. 3a. There reggeons can be replaced by scattering amplitudes
in the Regge asymptotics as in Fig. 3c.  Then the longitudinal momentum
integration can be performed easily by using the analytic properties of
the amplitudes $A $ and $B $ and substituting their dispersion
integrals.
Because of the cut-off condition (2.9) on the
heavy modes the virtual
gluon exchange in Fig. 3b does not correspond to $A $ and $B $ with
normal analytic
properties. Such properties hold, if we add Fig. 2 a,c,d.  This is the
way
how in the presence of the cut-off $q_0 $ one arrives at the box graph
including both the modes close and away from mass shell and obtains
reggeization of the exchanged gluon. The dependence of the longitudinal
momentum integral on the transverse momenta in the loop can be
disregarded in the leading log approximation. We obtain the result for
the sum of Fig. 2  by replacing in (2.7) the longitudinal momentum
integral as
\begin{eqnarray}
\int {dk_+ dk_- \over k_+ k_- } \longrightarrow
\int {s \ \ \ dk_+ dk_- \over
[(p_A - k_0 )^2 + i \epsilon ]
[(p_B + k_0 )^2 + i \epsilon ] }
\sim \ln { s \over q_0^2 }
\end{eqnarray}
$k_0 $ is the four-vector with the longitudinal momenta $k_+, k_- $ and
the transverse momenta replaced by $q_0$.

Generalizing this observation we conclude that graphs with triple
vertices of exchanged gluons can be regarded as reggeon diagrams, in
particular for doing the longitudinal momentum integrals, if one adds
appropriate contributions with with s-channel gluons close to mass
shell.

\section{The leading reggeon interaction}
\setcounter{equation}{0}

We consider the exchange of two gluons in the gauge group singlet
channel and their interaction. The BFKL \cite{BFKL} kernel of reggeized
gluon interaction is usually obtained by contracting two effective
vertices of gluon production ${\cal L}_p $ and including the gluon
reggeization. Alternatively it can be obtained from the triple vertex of
exchanged gluons (${\cal L}_t $) by interpreting the longitudinal
momentum integration in the way appropriate to reggeon diagrams. This is
the way how the derivation of the BFKL interaction kernel proposed in
\cite{AWCC1} fits in the framework of the QCD multi-Regge effective
action.

The kernel is obtained as the contribution from the intermediate state
in the $t$-channel with 3 exchanged gluons. The transition from 2 to 3
exchanged gluons is given by the graphs in Fig. 4  and by squaring we
obtain the graphs Fig. 5. In the gauge group singlet channel all these
graphs Fig. 5 give rise to the same gauge group factor. Also the
longitudinal
momentum integrals are  the same up to signs. In fact the graphs
contribute
to the $2\rightarrow 2 $ reggeon Green function $f$ in the scattering
amplitude Fig. 6. Therefore it is correct to evaluate the longitudinal
momentum integrals as above including at this step implicitely
contributions with intermediate scattering gluons close to mass shell.
As contributions to the amplitude Fig. 6  the longitudinal momentum
integral for each of the graphs Fig. 5  results in the same factor $\sim
\ln {s \over q_0^2 } $ in the leading log approximation.

The transverse momentum factors of the graphs give in the sum
\begin{equation}
{\vert \kappa_1 \vert^2 \vert \kappa_2^{\prime } \vert^2 \over
 \vert \kappa_1 - \kappa_1^{\prime } \vert^2 } +
{\vert \kappa_2 \vert^2 \vert \kappa_1^{\prime } \vert^2 \over
 \vert \kappa_1 - \kappa_1^{\prime } \vert^2 } -
\vert \kappa_2 \vert^2  \alpha (\kappa_1 )  \delta (\kappa_2 -
\kappa_2^{\prime })
- \vert \kappa_1 \vert^2  \alpha (\kappa_2 )  \delta (\kappa_1 -
\kappa_1^{\prime }),
\end{equation}
where the terms are ordered as the graphs in Fig. 5  and the
delta-function of momentum conservation $\kappa_1 + \kappa_2 =
\kappa_1^{\prime } + \kappa_2^{\prime } $ has been omitted.
$\alpha (\kappa ) $ denotes the transverse momentum integral in
(2.8) which determines the gluon trajectory.
The minus sign in front of the last two terms arises from the
logitudinal momenta in the vertices (2.6)
The weight ${1 \over 2} $ in the trajectory integral $\alpha (k) $ is
due to Bose symmetry.

By evaluating the longitudinal intergral as in reggeon diagrams we
include contributions beyond the ones originally represented by the
graphs with the triple vertices of exchanged gluons (2.6). Following
this line we cannot be sure that we recover all leading contributions.
Instead of going back to the effective action and analyzing all
contributions one can use at this point the property of reggeon Green
functions to vanish with vanishing transverse momenta.
Whereas the last two terms in (3.1) satisfy this condition trivially the
first two terms do not. Obviously the term
\begin{equation}
- \vert \kappa_1 + \kappa_2 \vert^2
\end{equation}
has to be added to (3.1). The missing contribution corresponds to a new
quartic vertex of exchanged gluons $({\cal A_- A_- A_+ A_+}) $
 Fig. 7 having the same gauge group factor and longitudinal
momentum
integrals as the graphs Fig. 5  but the transverse momentum structure
(3.2).    The sum of (3.1) and (3.2)
is the known BFKL interaction kernel.  Its connected part (omitting
the last two terms in (3.1) ) is denoted by $K_{2 \rightarrow 2} $
according to \cite{JB},
\begin{eqnarray}
K_{2 \rightarrow 2} (\kappa_1, \kappa_2; \kappa_1^{\prime },
\kappa_2^{\prime } ) =
- \vert \kappa_1 + \kappa_2 \vert^2 +
{\vert \kappa_1 \vert^2 \vert \kappa_2^{\prime } \vert^2
\ + \ \vert \kappa_2 \vert^2 \vert \kappa_1^{\prime } \vert^2  \over
\vert \kappa_1 - \kappa_1^{\prime } \vert^2 }
\cr =
{\kappa_1 \kappa_1^{\prime *}  \kappa_2^* \kappa_2^{\prime }
\ + \ {\rm c.c. } \over
\vert \kappa_1 - \kappa_1^{\prime } \vert^2 }.
\end{eqnarray}
The second form correspond to the transverse momentum part of the graph
Fig. 8a, obtained by contracting two effective production vertices
${\cal L}_p $, and the two terms in the numerator correspond to the two
helicities of the s-channel gluon.

The transition kernels $K_{2 \rightarrow 3}, K_{2 \rightarrow 4 }$,
which are a result of the s-channel unitarity ana\-ly\-sis in \cite{JB}
and which are building
blocks of the $2 \rightarrow 4 $ reggeon Green function \cite{AWCC1},
can
be most easily obtained as the transverse momentum factors of the graphs
Fig. 9a and 10a, correspondingly. With the effective vertices of
scattering ${\cal L}_s $ and production ${\cal L}_p $ we obtain
\begin{eqnarray}
K_{2 \rightarrow 3} (\kappa_1, \kappa_2 ; \kappa_1^{\prime },
\kappa_2^{\prime }, \kappa_3^{\prime } ) =
{\kappa_1 \kappa_1^{\prime *}  \kappa_2^* \kappa_3^{\prime }
 \over
( \kappa_1 - \kappa_1^{\prime })^* (\kappa_2 - \kappa_3^{\prime }) }
+ {\rm c.c. } \ \ \ ,
\cr
K_{2 \rightarrow 4} (\kappa_1, \kappa_2 ; \kappa_1^{\prime },
\kappa_2^{\prime }, \kappa_3^{\prime }, \kappa_4^{\prime } ) =
{\kappa_1 \kappa_1^{\prime *}  \kappa_2^* \kappa_4^{\prime }
 \over
( \kappa_1 - \kappa_1^{\prime })^* (\kappa_2 - \kappa_4^{\prime }) }
+ {\rm c.c. } \ \ \ .
\end{eqnarray}

In \cite{JB93} all t-channel gauge group and signature states of the two
incoming and the four outgoing reggeized gluons in the $2 \rightarrow 4
$ reggeon Green function are considered. Clearly this variety of cases
cannot be reconstructed by the vertex (1.1) and therefore the reggeon
Green
functions in \cite{AWCC1} and \cite{JB93} are not compatible. However
the following discussion  shows, that also here a t-channel approch
works. We rederive $K_{2 \rightarrow 3}, K_{2 \rightarrow 4 }$, using
$K_{2 \rightarrow 2}$ (3.3), which we have already obtained by
t-channel
analysis, and the triple vertex of exchanged gluons (2.6). We disregard
now the gauge group structure and concentrate on the transverse momentum
factors corresponding to the graphs only.

Consider the graphs in Fig. 11. They are obtained by contracting the
elementary splitting $2 \rightarrow 3$, Fig. 4, with a pairwise
interaction of the 3 exchanged gluons via $K_{2 \rightarrow 2}$. The
right-hand side is obtained by inserting
$K_{2 \rightarrow 2}$ as the sum of graphs Fig. 8b, where for
convenience we write the minus sign associated above with the quartic
vertex Fig. 7 (3.2) explicitely in front of the graph. Further we have
contracted those lines the propagators of which are cancelled by factors
in the adjacent vertices. Representing now merely transverse momentum
expressions, the meaning of the graphs, in particular of the vertex
(${\cal A_-  A_+  A_+  A_+ }$) emerging by this contraction, is obvious.

Consider Fig. 12, where now the incoming two gluons interact via $K_{2
\rightarrow 2}$ and the splitting Fig. 4  appears afterwards. Comparing
the resulting graphs   we observe, that the difference of Fig. 11 and
Fig. 12
is $ 2 \ K_{2 \rightarrow 3} $, where $ K_{2 \rightarrow 3}$ is given by
the sum of graphs in Fig. 9b  or by the expression
(with the terms in the same order as the graphs)
\begin{eqnarray}
K_{2 \rightarrow 3} (\kappa_1, \kappa_2 ; \kappa_1^{\prime },
\kappa_2^{\prime }, \kappa_3^{\prime } ) =
- \vert \kappa_1 + \kappa_2 \vert^2 +
{ \vert \kappa_1^{\prime } + \kappa_2^{\prime } \vert^2 \vert \kappa_2
\vert^2 \over
\vert \kappa_2 - \kappa_3^{\prime } \vert^2 }
+ { \vert \kappa_2^{\prime } + \kappa_3^{\prime } \vert^2 \vert \kappa_1
\vert^2 \over
\vert \kappa_1 - \kappa_1^{\prime } \vert^2 }
\cr
- {\vert \kappa_1 \vert^2 \vert \kappa_2 \vert^2 \vert \kappa_2^{\prime
} \vert^2 \over
\vert \kappa_2 - \kappa_3^{\prime } \vert^2
\vert \kappa_1 - \kappa_1^{\prime } \vert^2 } .
\end{eqnarray}
We have seen that
Bartels' transition kernel $K_{2 \rightarrow 3} $ is proportional to the
connected part of the convolution of the elementary splitting $2
\rightarrow 3 $ with the pairwise interaction via $K_{2 \rightarrow 2}$
modulo $K_{2 \rightarrow 2}$ convoluted with subsequent $2 \rightarrow
2$ elementary splitting.

The analogous relation holds for the transition
kernel $K_{2 \rightarrow 4}$, only the interaction $K_{2 \rightarrow 2}$
is to be replaced by the transition of pairs the exchenged gluons into
three gluons via $K_{2 \rightarrow 3}$.
Indeed, the difference of Fig. 13  and Fig. 14  is $ 3 \ K_{ 2
\rightarrow
4}$, where $K_{ 2 \rightarrow 4}$ is represented by the sum of graphs in
Fig. 10b  or by the sum of terms corresponding to these 4 graphs,
\begin{eqnarray}
K_{2 \rightarrow 4} (\kappa_1, \kappa_2 ; \kappa_1^{\prime },
\kappa_2^{\prime }, \kappa_3^{\prime }, \kappa_4^{\prime } ) =
- \vert \kappa_1 + \kappa_2 \vert^2 +
{ \vert \kappa_1^{\prime } + \kappa_2^{\prime } +
\kappa_3^{\prime } \vert^2 \vert \kappa_2 \vert^2 \over
\vert \kappa_2 - \kappa_4^{\prime } \vert^2 }
\cr
+ { \vert \kappa_2^{\prime } + \kappa_3^{\prime } +
\kappa_4^{\prime } \vert^2 \vert \kappa_1 \vert^2 \over
\vert \kappa_1 - \kappa_1^{\prime } \vert^2 }
- {\vert \kappa_1 \vert^2 \vert \kappa_2 \vert^2 \vert \kappa_2^{\prime
} + \kappa_3^{\prime } \vert^2 \over
\vert \kappa_2 - \kappa_4^{\prime } \vert^2
\vert \kappa_1 - \kappa_1^{\prime } \vert^2 } .
\end{eqnarray}

The coincindence of the expressions (3.4) , (3.5) for $K_{2
\rightarrow
3}$ and (3.4), (3.6) for $K_{2 \rightarrow 4}$ can be checked by
straightforward calculations. We recommend to consider the limiting
cases of vanishing and large momentum transfer $\kappa_1 + \kappa_2 $.

\section{The one-loop  reggeon interaction}
\setcounter{equation}{0}

The one-loop corrections arise from the s-channel configurations, where
pairs of particles do not have a large sub-energy and their longitudinal
momenta violate the multi-Regge strong ordering. It is not obvious at
all that the procedure, where longitudinal momentum integrals are
treated as in reggeon diagrams, can give some reasonable result here.
In the $t$-channel approach the one-loop correction to the BFKL kernel
arises from the contribution to the two-gluon exchange (Fig. 6) by
 4 gluons in the $t$-channel intermediate state. Analogous to the
case of 3 intermediate gluons above the integration over the
longitudinal momenta has to be considered for the graphs inserted in the
amplitude Fig. 6. But now only one of the two loop integrals with the
additional intermediate gluons gives a contribution $ \sim \ln s$. The
leading log arguments which allowed above to separate the logitudinal
from the transverse momentum intergrals are not applicable here.

Our point is to formulate assumptions under which such a separation
would hold and under which the formula obtained in \cite{AWCC1} would
make sense.
Unavoidably this takes to introduce a scale, which cannot be fixed
within the framework of the approach.

At this place one might feel disappointed and  prefer to wait for
the
full QCD next-to-leading calculation. Nevertheless it may be worthwile
to look for a scheme allowing eventually to obtain a simple estimate of
the higher loop corrections.

We discuss schematically the integration over the longitudinal momenta.
Consider the contributions to the amplitude Fig. 15.
Treating the exchanged gluons as reggeons the longitudinal momentum
integrals over $k_{2 +}, k_{1 +}$ can be  done by taking residues
(under the dispersion integrals for the amplitudes $A,  B $) at
appropriate poles. We arrive in the second case, Fig. 15 b, at
\begin{eqnarray}
\int_{q_0}^{\sqrt s} {dk_{1 -} dk_{2 -} \over k_{1 -} k_{2 -} }
\theta (k_{1 -} - k_{2 -} )
= \int_0^{Y} dy_{1 -} dy_{2 -} \theta (y_{1 -} - y_{2 -})
= \frac{1}{2} Y^2.
\end{eqnarray}
We use the notations $y_{-} = \ln {k_- \over q_0}, Y = \frac{1}{2} \ln
{s \over q_0^2} $.  In the first case, Fig. 15 a,  we encounter
contradicting ordering
conditions. The result is complicated, depending in particular on the
transverse momenta.

In order to proceed without working out the details, one has to assume
that these intergals can be approximated nevertheless by logarithmic
integrals in $k_{1 -}, k_{2 -} $ with relaxed ordering conditions:
\begin{equation}
\int_0^{Y} dy_{1 -} dy_{2 -} \theta (y_{1 -} - y_{2 -} + \eta )
 \theta (y_{2 -} - y_{1 -} + \eta )
= Y \ 2 \eta.
\end{equation}
In any case, at large $Y$, the result can be written in this form, but
then $\eta $ is a function depending on transverse momenta and this
dependence is different for each graph Fig. 17.

One can write $\eta = \ln {q_1 \over q_0 }$ and interprete $q_0$ as the
typical transverse momentum and $q_1$ as characterizing the width of the
transverse momentum range.

We consider the splitting of  two exchanged gluons in the gauge group
singlet state into 4 gluons, Fig. 16. The gauge group state of the 4
gluons can be characterized by the quantum numbers in a two-gluon
sub-channel as in \cite{JB93}. One can choose the two gluons the
$k_+$ of which enters the expression. We give them the numbers 2 and 3.
Only the state symmetric in the gauge group indices of these two gluons
contributes as an intermediate state to the kernel. The antisymmetric
(adjoint representation) state contributes to reggeization and will be
accounted
for elsewhere. The longitudinal momenta of these two gluons are close to
each other,
\begin{equation}
\vert \ln {k_{2 +} \over k_{1 +} } \vert < \eta ,   \ \ \ \
\vert \ln {k_{2 -} \over k_{1 -} } \vert < \eta .
\end{equation}

We obtain the correction to the kernel by contracting the $2 \rightarrow
4 $ graphs Fig. 16 with corresponding $4 \rightarrow 2$ graphs. The
longitudinal momentum intergals, considered as contributions to the
amplitude Fig. 6 and approximated as in (4.2),  are universal in all
terms. Also the gauge group matrices give rise to an universal factor
proportional to
\begin{equation}
\left ( {\bf tr} (T^{a} T^{b} T^{a} T^{b} ) +
 {\bf tr} (T^{a} T^{b} T^{b} T^{a} ) \right ) / (N^2 - 1).
\end{equation}
After the separation of the longitudinal momentum and the gauge group
structures the graphs can be considered as representing transverse
momentum expressions only. As in the previous section we contract the
intermediate lines, e.g. in Fig. 16a and 16b,  the propagators of which
are cancelled by factors in the vertices.
We arrive at the graphs in Fig. 17. As a remainder from the
longitudinal
momenta the graphs in the second line enter with a minus sign.
In front of the loop integrals the following Bose symmatry factors
appear: ${1 \over 3 !} $ in the first two graphs, $({1 \over 2 !})^2$ in
the
third graph and ${1 \over 2 !}$ in the remaining graphs besides of the
last one.
 Since the expression can be read off easily from
the graphs we shall not write it here. We check that the resulting
one-loop correction to the reggeon Green function obeys the conditon of
vanishing with any of the transverse momenta. No new vertex and no loop
correction in the existing vertices has to be introduced.

Summarizing,  we have seen that in order to reproduce the result
of \cite{AWCC1} the longitudinal momentum integration has to be
approximated and the gauge group intermendiate state has to be
restricted. Therefore the one-loop correction enters the reggeon
interaction kernel with the factor
\begin{equation}
\eta{3 \over 2 } N^2 .
\end{equation}

It will be interesting to compare with the full QCD calculation and to
ask whether it can be represented by transverse momentum expressions
corresponding to the graphs Fig. 17 with each term modified by
a function $\eta (\kappa_1
...)$ inserted in the integrand, weakly depending on the transverse
momenta and different from term to term.

\section{Calculating the eigenvalues}
\setcounter{equation}{0}

Here we outline our method of the eigenvalue calculation for the
one-loop interaction kernel given by the graphs Fig. 17 in the forward
limit, $\kappa_2 = - \kappa_1 = \kappa,
\kappa_2^{\prime } = - \kappa_1^{\prime } = \kappa^{\prime } $.
\begin{equation}
\int { d^2\kappa^{\prime } \over \vert \kappa^{\prime } \vert^4 }
K^{(4)} (\kappa, \kappa^{\prime } ) f_{n, \nu }(\kappa^{\prime })
= \Omega^{(4)} (n, \nu ) f_{n, \nu } (\kappa ).
\end{equation}
The eigenfunctions are the same as for the lowest order kernel
\cite{Lev86},
\begin{equation}
f_{n, \nu }(\kappa ) = \vert \kappa^2 \vert^{1/2 + i \nu }
\ \ \left ( { \kappa \over \vert \kappa \vert } \right )^n
\end{equation}
($-\infty < \nu < \infty , n $ integer ),
and the one-loop kernel at vasnishing momentum transfer is given by
\begin{eqnarray}
K^{(4)} (\kappa ,\kappa^{\prime } ) =
(2 \pi )^3 \vert \kappa \vert^6 \left (\frac{1}{3} J_2 (\kappa) +
\frac{1}{4} (J_1 (\kappa ) )^2 \right ) \delta (\kappa - \kappa^{\prime
} ) \cr
- \vert \kappa - \kappa^{\prime } \vert^{-2}
\left ( 2 \vert \kappa \vert^4 J_1 (\kappa ) \vert \kappa^{\prime }
\vert^2 \ + \ 2 \vert \kappa \vert^2 J_1 (\kappa^{\prime } ) \vert
\kappa^{\prime } \vert^4  \right )
\cr
+ \ 2 J_1 (\kappa - \kappa^{\prime } ) \vert \kappa \vert^2 \vert
\kappa^{\prime } \vert^2 +
2 \vert \kappa \vert^4 \ \vert \kappa^{\prime } \vert^4
I(\kappa ,\kappa^{\prime })
\cr
+ (\kappa^{\prime }  \leftrightarrow - \kappa^{\prime } )
\end{eqnarray}
The terms are ordered in the same way as the graphs in Fig. 17.  By
adding the corresponding expression with the opposite sign in front of
$\kappa^{\prime } $ one projects on the positive signature channel. The
notation for the loop integrals are the same as in \cite{AWCC1},
\begin{eqnarray}
J_1 (\kappa ) = (2 \pi )^{-3 } \int d^2\kappa^{\prime }
\vert \kappa^{\prime } \vert^{-2}
\vert \kappa - \kappa^{\prime } \vert^{-2}
= \vert \kappa \vert^{-2} \alpha ( \kappa ),
\cr
J_2 (\kappa ) = (2 \pi )^{-3 } \int d^2\kappa^{\prime }
\vert \kappa - \kappa^{\prime } \vert^{-2}
J_1 (\kappa^{\prime } ),
\cr
I (\kappa, \kappa^{\prime } ) = (2 \pi )^{-3 } \int
d^2 \kappa^{\prime \prime}
\vert \kappa^{\prime \prime } ( \kappa^{\prime \prime } - \kappa )
 ( \kappa^{\prime \prime } + \kappa^{\prime } )
( \kappa^{\prime \prime } - \kappa + \kappa^{\prime } ) \vert^{-2} .
\end{eqnarray}
The integrands are moduli squared of rational expressions. The latter
can be decomposed into simple fractions. This is the essential step in
calculating the contribution of $I(\kappa , \kappa^{\prime })$ to the
eigenvalue in our approach.
\begin{eqnarray}
[ \kappa^{\prime \prime } ( \kappa^{\prime \prime } - \kappa )
 ( \kappa^{\prime \prime } + \kappa^{\prime } )
( \kappa^{\prime \prime } - \kappa + \kappa^{\prime } ) ]^{-1}
={ 1 \over \kappa \kappa^{\prime } }
\lbrace {1 \over \kappa^{\prime } - \kappa }
\left ( {1 \over \kappa^{\prime \prime } } -
{1 \over \kappa^{\prime \prime } - \kappa + \kappa^{\prime } }
\right )
\cr
 + {1 \over \kappa^{\prime } + \kappa }
\left ( {1 \over \kappa^{\prime \prime } - \kappa } -
{1 \over \kappa^{\prime \prime }  + \kappa^{\prime } }
\right )  \rbrace
\end{eqnarray}
In this way $I(\kappa, \kappa^{\prime } )$ is decomposed into two parts,
\begin{eqnarray}
2 \vert \kappa \vert^4 \ \vert \kappa^{\prime } \vert^4
I(\kappa ,\kappa^{\prime })  +
(\kappa^{\prime }  \leftrightarrow - \kappa^{\prime } )
= K^{(4)}_2 (\kappa ,\kappa^{\prime })
\cr
+ {2 \over (2 \pi )^3 } \int d^2 \kappa^{\prime \prime }
\vert {1 \over \kappa^{\prime \prime }  } -
{1 \over \kappa^{\prime \prime }  - \kappa + \kappa^{\prime } }
\vert^2 \vert \kappa^{\prime } - \kappa \vert^{-2}
\vert \kappa \vert^2 \vert \kappa^{\prime } \vert^2
+ (\kappa^{\prime }  \leftrightarrow - \kappa^{\prime } ),
\end{eqnarray}
where
\begin{eqnarray}
 K^{(4)}_2 (\kappa ,\kappa^{\prime })
=
 {2 \over (2 \pi )^3 } \int d^2 \kappa^{\prime \prime }
 ( {1 \over \kappa^{\prime \prime } } -
{1 \over \kappa^{\prime \prime } - \kappa + \kappa^{\prime } } )
 ( {1 \over \kappa^{\prime \prime } - \kappa } -
{1 \over \kappa^{\prime \prime }  + \kappa^{\prime } } )^*
\cr
\cdot ( \kappa^{\prime } - \kappa )^{-1}
( \kappa^{\prime } + \kappa )^{* -1}
+ {\rm c.c. }
\end{eqnarray}
Because the integral in the second term of (5.6) is $J_1 (\kappa -
\kappa^{\prime }) \vert \kappa - \kappa^{\prime } \vert^2 $, this term
coincides with the last of the remaining terms in $K^{(4)}$ (5.3). It
is natural to decompose the kernel,
\begin{equation}
K^{(4)} = K^{(4)}_1 + K^{(4)}_2.
\end{equation}
Moreover this decomposition is reasonalble owing to the infrared
divergence cancellation. The integral defining $K^{(4)}_2 $ and also the
intergral defining the action of $K^{(4)}_2 $ are free of divergencies.

It is not difficult to check that the divergencies in the terms of
$K^{(4)}_1 \otimes f $ cancel against each other. If we want to
calculate the eigenvalue of $K^{(4)}_1 $ term by term we have to
introduce a regularization. Because the decomposition (5.6) related to
the partial fractions works only in exactly 2 dimensions we avoid the
dimensional regularization. In the complex planes of $\kappa^{\prime } $
and $\kappa^{\prime \prime} $ we cut out  small discs of radius
$\lambda $ around all singularities. We represent the step function
imposing this cut-off by ($\delta \rightarrow +0 $)
\begin{equation}
\theta ( \vert \kappa^{\prime } - \kappa \vert - \lambda ) =
{1 \over 2 \pi i } \int_{- i \infty + \delta }^{i \infty + \delta }
{d\omega \over \omega }
\left ( {\vert \kappa^{\prime } - \kappa \vert^2 \over \lambda^2 }
\right )^{\omega }.
\end{equation}
With this representation the next steps of the calculation are close to
the ones in dimensional regularization. In the Appendix we explain the
calculation of $J_1 (\kappa ) $ and of the eigenvalue of one term in
$K^{(4)}_1 $. The resulting eigenvalue is \cite{AWCC1}
\begin{eqnarray}
\Omega^{(4)}_1 (n, \nu ) = \frac{2}{\pi } (1 + (-1)^n )
(\chi (n, \nu ))^2,
\cr
\chi (n, \nu ) =  \psi (1)
- \frac{1}{2} \psi ({1 \over 2} + i \nu + { \vert n \vert \over 2 } )
- \frac{1}{2} \psi ({1 \over 2} - i \nu + { \vert n \vert \over 2 } ).
\end{eqnarray}

Now we turn to $K^{(4)}_2 $ (5.7). We calculate the intergral over
$\kappa^{\prime \prime } $ reducing it by shifts to $J_1 (\kappa ) -
J_1 (\kappa^{\prime } ) $. This leads to the representation
\begin{eqnarray}
(2 \pi )^3 \int {d^2 \kappa^{\prime } \over \vert \kappa^{\prime }
\vert^4 }
K^{(4)}_2 (\kappa , \kappa^{\prime } ) f_{n, \nu } (\kappa^{\prime } )
= - 4 \pi f_{n, \nu } \ \ \
{d \over d\omega } J_{\omega }(n, \nu ) \ \vert_{\omega = 0 },
\cr
J_{\omega }(n, \nu ) = \int d^2 y
\vert y^2 \vert^{-1/2 + i \nu - \omega }
\left ( {y \over \vert y \vert } \right )^n
\left ( {1 \over (y - 1)(y + 1)^* } + {\rm c.c.} \right ).
\end{eqnarray}
We introduce polar coordinates, $y = r e^{i \phi } $, and do the
integral
over $\phi $ by taking residues in the complex plane of $z = e^{i \phi
}$.  In the intermedialte steps the $r$ integral is divided into $0 < r
< 1$ and $1 < r < \infty $. Inversion in the second term leads to
\begin{eqnarray}
J_{\omega }(n, \nu ) = 2 \pi (1 + (-1)^n ) \int_0^1
{d r \over r^2 + 1} r^{\vert n \vert }
(r^{-2 i \nu + 2 \omega } -
r^{2 i \nu - 2 \omega } ) .
\end{eqnarray}
This integral can now be expressed in terms of the incomplete
beta function \cite{GR},
\begin{equation}
\beta (p) = \int_0^1 {x^{p-1} dx \over x + 1 }
= \frac{1}{2} \left ( \psi ({p + 1 \over 2} ) - \psi ({p \over 2})
\right ).
\end{equation}
We obtain finally
\begin{equation}
\Omega^{(4)}_2 (n, \nu) = - {1 \over 2 \pi } (1 + (-1)^{n} ) \left (
\beta^{\prime } ({\vert n \vert + 1 \over 2 } + i \nu )
+ {\rm c.c.} \right ) ,
\end{equation}
where the prime denotes the derivative of the function.

The discussion of the interesting properties of the result related to
conformal symmetry and holomorphic factorization \cite{L90}
\cite{RK94} and the numerical
evaluation of the leading eigenvalue has been given in \cite{AWCC2}. We
emphasize that as a result of our analysis the one-loop kernel $K^{(4)}$
and therefore the eigenvalues $\Omega^{(4)} $ enter multiplied by the
factor (4.5).

\section{Discussion}

The multi-Regge effective action is a tool for investigating the Regge
asymptotics. The leading logarithmic approximation to the high-energy
asymptotics is reproduced easily by using the lowest order effective
vertices of scattering and production. The lowest order effective action
provides a starting point to study those corrections to the leading log
approximation, which are related to the multiple exchange of reggeized
gluons and which are responsible for restoring unitarity in all
sub-energy channels. Mostly this study has been approached from the
point of view of $s$-channel intermediate states. Alternatively the
point of view of $t$-channel intermediate states can be taken. It leads
to an essentially different way for obtaining the leading log amplitudes
and the unitarity corrections.

The gluons exchanged in the $t$-channel behave like signature-odd
reggeons. This observation is confirmed in particular by the similarity
of the triple vertex of exchanged gluons in the multi-Regge effective
action to the triple vertex of signature-odd reggeons considered in
\cite{ARW} \cite{AWCC1}. We have shown that starting from this vertex in
the effective action a $t$-channel approach similar to the one in
\cite{AWCC1} works. The BFKL leading log interaction kernel of reggeized
gluons and the transition kernels \cite{JB} \cite{JB93}, which are
relevant for the unitarity corrections, have been reconstructed in this
approach.

Graphs with the triple vertex of exchange gluons can be treated as
reggeon diagrams, in particular  the longitudinal momentum
integrals can be interpreted in this way. This amounts to include
implicitely contributions with scattering gluons in $s$-channel
intermediate states.

There are more corrections to the leading log approximation besides of
those which improve the unitarity properties. In the effective action
such next-to-leading log corrections give rise to corrections to the
effective vertices, to corrections to the trajectory and to new vertices
\cite{FL} \cite{Lev95}. The corrected production vertices and the
corrected trajectory result in the corrected BFKL kernel.

Starting from the effective action on the leading log level, in
particular from the triple vertex of exchanged gluons, and applying the
$t$-channel analysis following basically the one by White \cite{AWCC1},
we obtain an one-loop interaction kernel. Our discussion above shows to
what extend this kernel can be a reasonable estimate of the full
next-to-leading
 log correction to the BFKL kernel. In particular we point out a
factor in front of this one-loop kernel involving a scale, which arises
from approximating the longitudinal momentum integrals. We expect that
the full next-to-leading log kernel has a similar structure with this
rapidity factor replaced in each term by a function weakly depending
on the transverse momenta.

The loop integrals encountered in the eigenvalue calculation can be done
without major effort by using the simplicity of the integrands. They are
moduli squared of rational expressions, which can be decomposed into
simple fractions.

\newpage

$\mbox{ }$ \\
{\Large\bf Acknowledgements} \\
$\mbox{ }$ \\
The author is grateful to A.R. White for discussions and for
hospitality at Argonne. He had also useful discussions with J. Bartels,
L.N. Lipatov and L. Szymanowski.
The financial support by Deutsche Forschungsgemeinschaft is gratefully
acknowledged.

\vspace{1cm}

$\mbox{ }$ \\
{\Large\bf Appendix} \\
$\mbox{ }$ \\

We calculate $J_1 (\kappa )$ (5.4) regularized as described (5.9).
\begin{eqnarray}
(2 \pi )^3 J_1 (\kappa ) \vert_{reg} =
\int {d^2 \kappa^{\prime } \over
\vert \kappa^{\prime } \vert^{2}
\vert \kappa - \kappa^{\prime } \vert^{2} }
\theta ( \vert \kappa^{\prime } \vert - \lambda )
\theta ( \vert \kappa^{\prime } - \kappa \vert - \lambda )
\cr
= {1 \over (2 \pi i )^2 } \int {d\omega_0 \  d\omega  \over \omega_0 \
\omega }  \lambda^{ -2\omega_0 - 2\omega }
\int {d^2 \kappa^{\prime } \over
\vert \kappa^{\prime } \vert^{2-2\omega_0}
\vert \kappa - \kappa^{\prime } \vert^{2-2 \omega } }.
\end{eqnarray}
We do the integral over $\kappa^{\prime }$ and arrive at
\begin{eqnarray}
 {\pi \over (2 \pi i )^2 } \int {d\omega_0 \  d\omega  \over \omega_0 \
\omega }  \left ( {\vert \kappa \vert^2 \over \lambda^2 } \right )^{
-\omega_0 - \omega}
\left ({1 \over \omega_0 \ } + {1 \over \omega } \right )
\cr
{ \Gamma (1- \omega_0 - \omega ) \Gamma (1 + \omega ) \Gamma (1 +
\omega_0 ) \over
 \Gamma (1+ \omega_0 + \omega ) \Gamma (1 - \omega ) \Gamma (1 -
\omega_0 ) }.
\end{eqnarray}
For $\vert \kappa \vert > \lambda $ we obtain
\begin{equation}
 {\pi \over (2 \pi i ) } \int {  d\omega  \over \omega^2 }
\vert {\kappa^2 \over \lambda^2 } \vert^{\omega } =
2 \pi \ln \vert {\kappa^2 \over \lambda^2 } \vert.
\end{equation}

We pick up one term out of $K_1^{(4)} \otimes f_{n, \nu }$:
\begin{eqnarray}
(2 \pi )^3 \int d\kappa^{\prime }  J_1 (\kappa - \kappa^{\prime } )
\vert \kappa^{\prime } \vert^{-2} \vert \kappa \vert^{-2}
f_{n, \nu} (\kappa^{\prime } ) \ \ \vert_{reg}
\cr
= {1 \over (2 \pi i )^3 } \int {d\nu_0 \  d\nu_1 \ d\omega  \over \nu_0
\ \nu_1 \  \omega }  \lambda^{ -2\nu_0 -2\nu_1 - 2\omega }
2 \pi
\int {d^2 \kappa^{\prime } \kappa^{\prime n} \over
\vert \kappa^{\prime } \vert^{2-2 \nu_0 - 1 -2i \nu}
\vert \kappa - \kappa^{\prime } \vert^{2 - 2\nu_1- 2 \omega } }
\cr
= f_{n, \nu} (\kappa )
 {2 \pi^2 \over 2 \pi i  } \int {d\omega \over
\omega^3 }
  \left ( {\vert \kappa \vert^2 \over \lambda^2 } \right )^{
\omega}
{\Gamma ({1 \over 2} + i \nu + {\vert n \vert \over 2} )
\Gamma (1 + \omega )
\Gamma ({1 \over 2} - i \nu + {\vert n \vert \over 2} - \omega )
\over
\Gamma ({1 \over 2} + i \nu + {\vert n \vert \over 2} + \omega )
\Gamma (1 - \omega )
\Gamma ({1 \over 2} - i \nu + {\vert n \vert \over 2} )  }
\cr
= f_{n, \nu} (\kappa ) 2 \pi^2 \lbrace \frac{1}{2}
\ln^2 \vert {\kappa^2 \over \lambda^2 } \vert
+ \ln \vert {\kappa^2 \over \lambda^2 } \vert   \ 2 \chi (n, \nu )
+  {d \over d(i \nu) } \ \chi (n, \nu)  +
2 (\chi (n, \nu ) )^2 \rbrace .
\end{eqnarray}
We have used the notation as in (5.10).

\newpage
\noindent{\Large\bf Figure captions}
\vspace{2cm}

\begin{tabular}{ll}

Fig. 1  &  Vertices of the effective action. a) Effective vertex of \\
        &  emission, b) effective vertex of scattering,     \\
        &  c) triple vertex of exchanged gluons.         \\
Fig. 2  &  Contributions to reggeization of exchanged gluons.\\
Fig. 3  &  Interpretation of the longitudinal momentum integral. \\
Fig. 4  &  $2 \rightarrow 3$ splitting of exchanged gluons. \\
Fig. 5  &  Contributions to the $2 \rightarrow 2 $ Green function of
           reggeized gluons. \\
Fig. 6  &  Contribution of the $ 2 \rightarrow 2$ reggeon Green \\
        &  function  to the scattering amplitude. \\
Fig. 7  &  Quartic vertex of exchanged gluons. \\
Fig. 8  &  The transition kernel $2 \rightarrow 2$ \\
        &  in s-channel (a) and t-channel (b) representations. \\
Fig. 9  &  The transition kernel $2 \rightarrow 3$        \\
        &  in s-channel (a) and t-channel (b) representations.  \\
Fig. 10 &  The transition kernel $2 \rightarrow 4$  \\
        &  in s-channel (a) and t-channel (b) representations. \\
Fig. 11 &  Contraction of the $2 \rightarrow 3$ splitting with the
          $K_{2 \rightarrow 2}$ interaction.  \\
Fig. 12 &  Contraction of $K_{2 \rightarrow 2}$ with the
          $2 \rightarrow 3$ splitting.  \\
Fig. 13 &  Contraction of the $2 \rightarrow 3$ splitting with
          the transition kernel $K_{2 \rightarrow 3} $. \\
Fig. 14 &  Contraction of
          the transition kernel $K_{2 \rightarrow 3} $ with the
           $3 \rightarrow 4$ splitting. \\
Fig. 15 &  Contributions to the scattering amplitude. \\
Fig. 16 &  Graphs contributing to the $2 \rightarrow 4 $ splitting. \\
Fig. 17 &  Transverse momentum graphs for the one-loop correction \\
        &  to the $2 \rightarrow 2$ reggeon Green function. \\
        &  \\

\end{tabular}

\newpage

%\end{document}

\begin{figure}[thb]
\centering
\epsfig{ file = 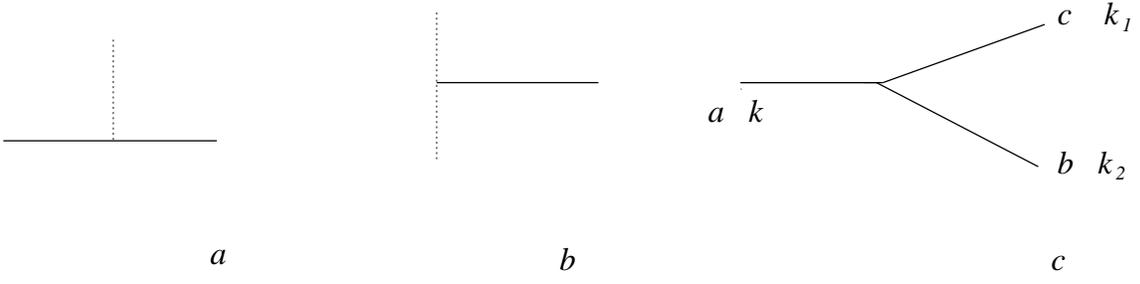, width = 150mm }
\caption{\em
  Vertices of the effective action. a) Effective vertex of
          emission, b) effective vertex of scattering,
          c) triple vertex of exchanged gluons.}
\end{figure}

\vspace{3cm}

\begin{figure}[thb]
\centering
\epsfig{ file = 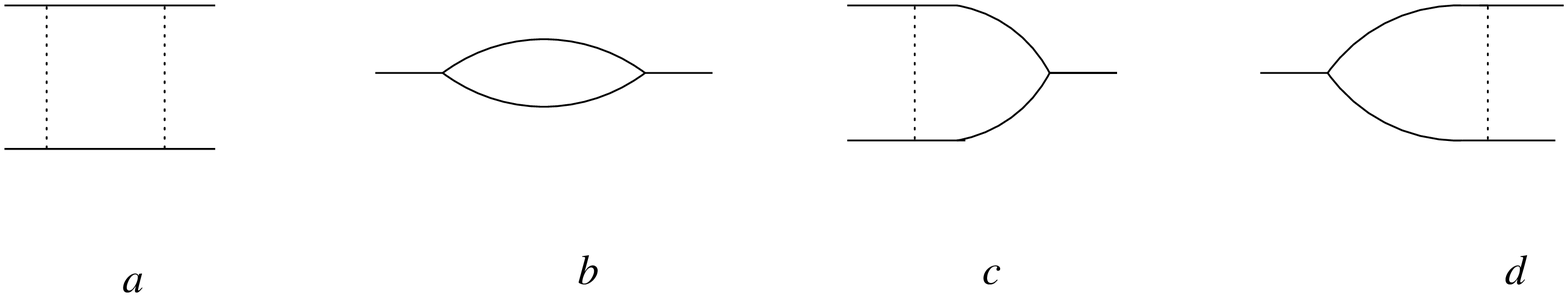, width = 150mm }
\caption{\em
  Contributions to reggeization of exchanged gluons.}
\end{figure}

\vspace{3cm}

\begin{figure}[thb]
\centering
\epsfig{ file = 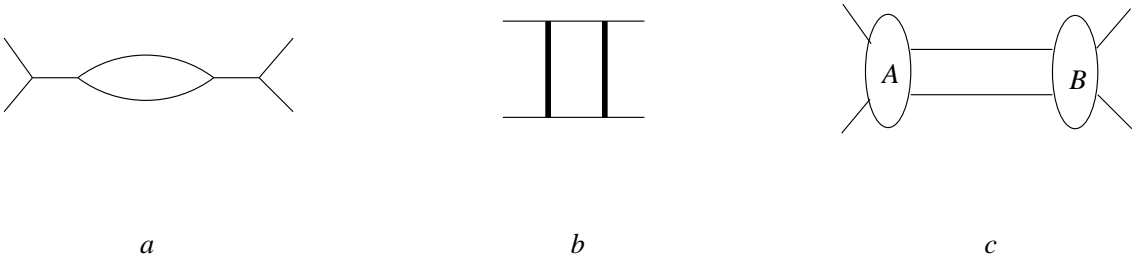, width = 150mm }
\caption{\em
  Interpretation of the longitudinal momentum integral.
           }
\end{figure}

\vspace{3cm}

\begin{figure}[thb]
\centering
\epsfig{ file = 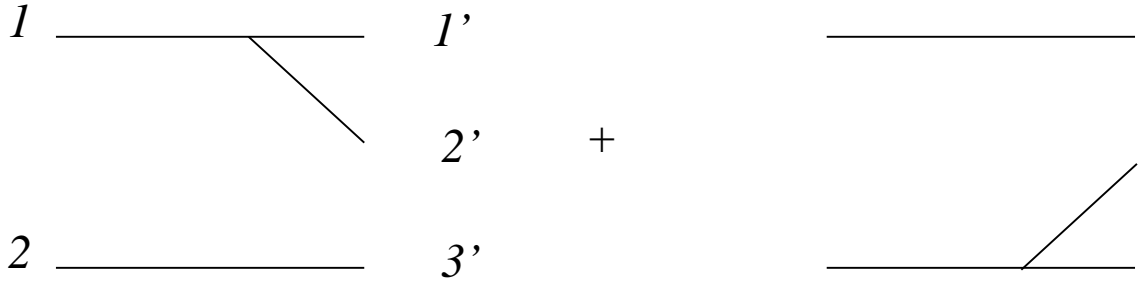, width = 150mm }
\caption{\em
  $2 \rightarrow 3$ splitting of exchanged gluons.}
\end{figure}

\vspace{3cm}

\begin{figure}[thb]
\centering
\epsfig{ file = 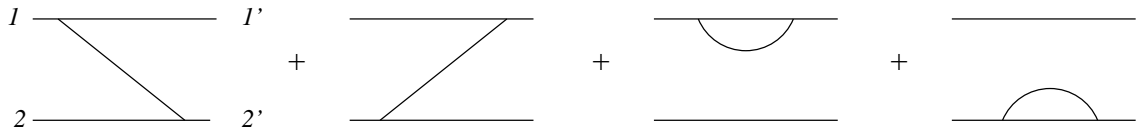, width = 150mm }
\caption{\em
  Contributions to the $2 \rightarrow 2 $ Green function of
           reggeized gluons.}
\end{figure}

\vspace{3cm}

\begin{figure}[thb]
\centering
\epsfig{ file = 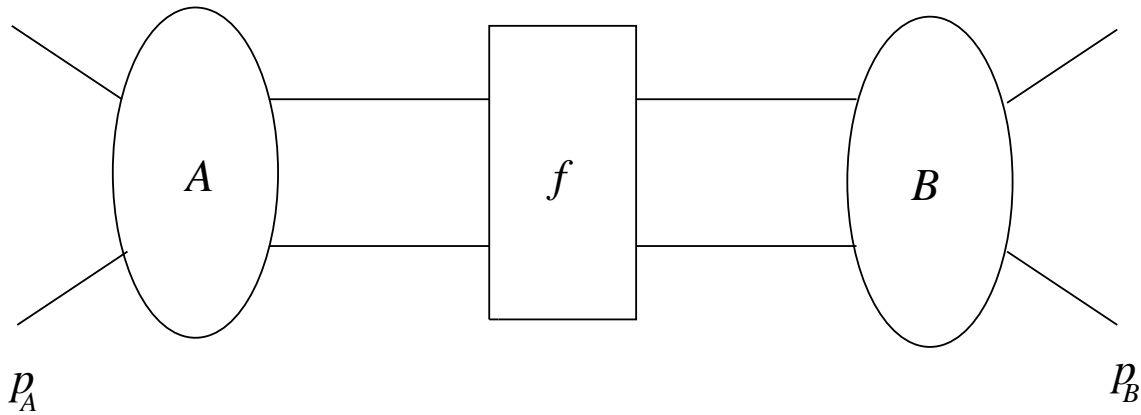, width = 150mm }
\caption{\em
  Contribution of the $ 2 \rightarrow 2$ reggeon Green
          function  to the scattering amplitude.}
\end{figure}

\vspace{3cm}

\begin{figure}[thb]
\centering
\epsfig{ file = 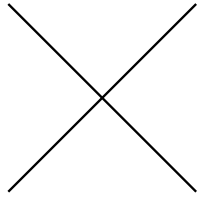, width = 25mm }
\caption{\em
  Quartic vertex of exchanged gluons. }
\end{figure}

\vspace{3cm}

\begin{figure}[thb]
\centering
\epsfig{ file = 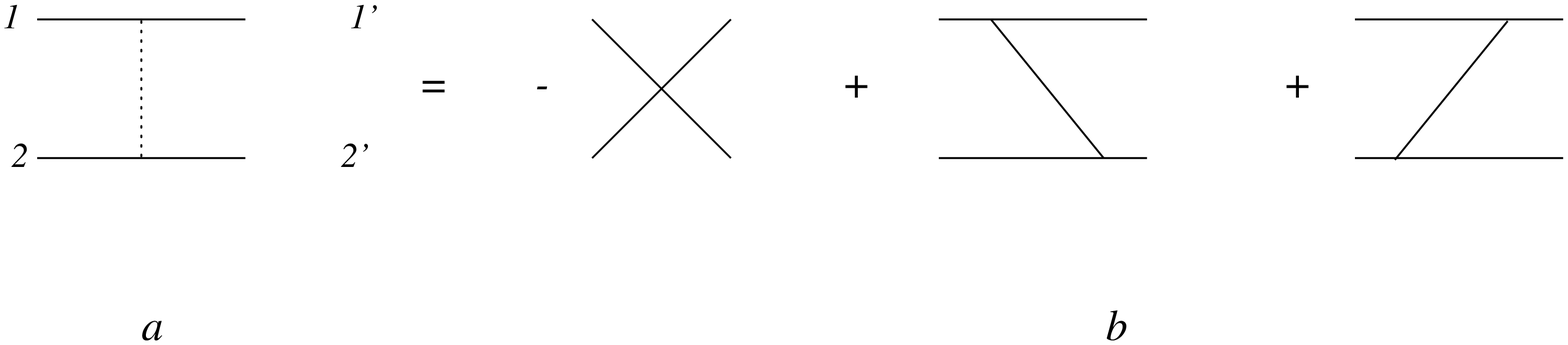, width = 150mm }
\caption{\em
  The transition kernel $2 \rightarrow 2$
          in s-channel (a) and t-channel (b) representations. }
\end{figure}

\vspace{3cm}

\begin{figure}[thb]
\centering
\epsfig{ file = 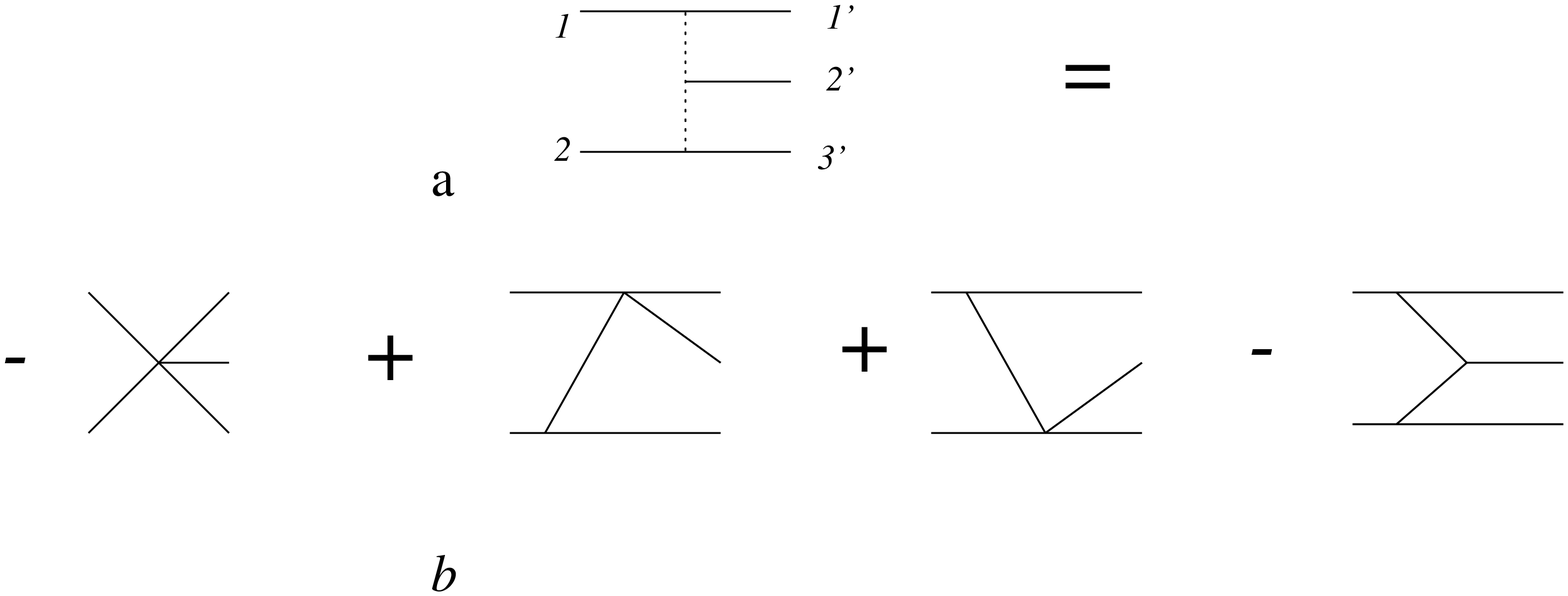, width = 150mm }
\caption{\em
 The transition kernel $2 \rightarrow 3$
          in s-channel (a) and t-channel (b) representations.  }
\end{figure}

\vspace{3cm}

\begin{figure}[thb]
\centering
\epsfig{ file = 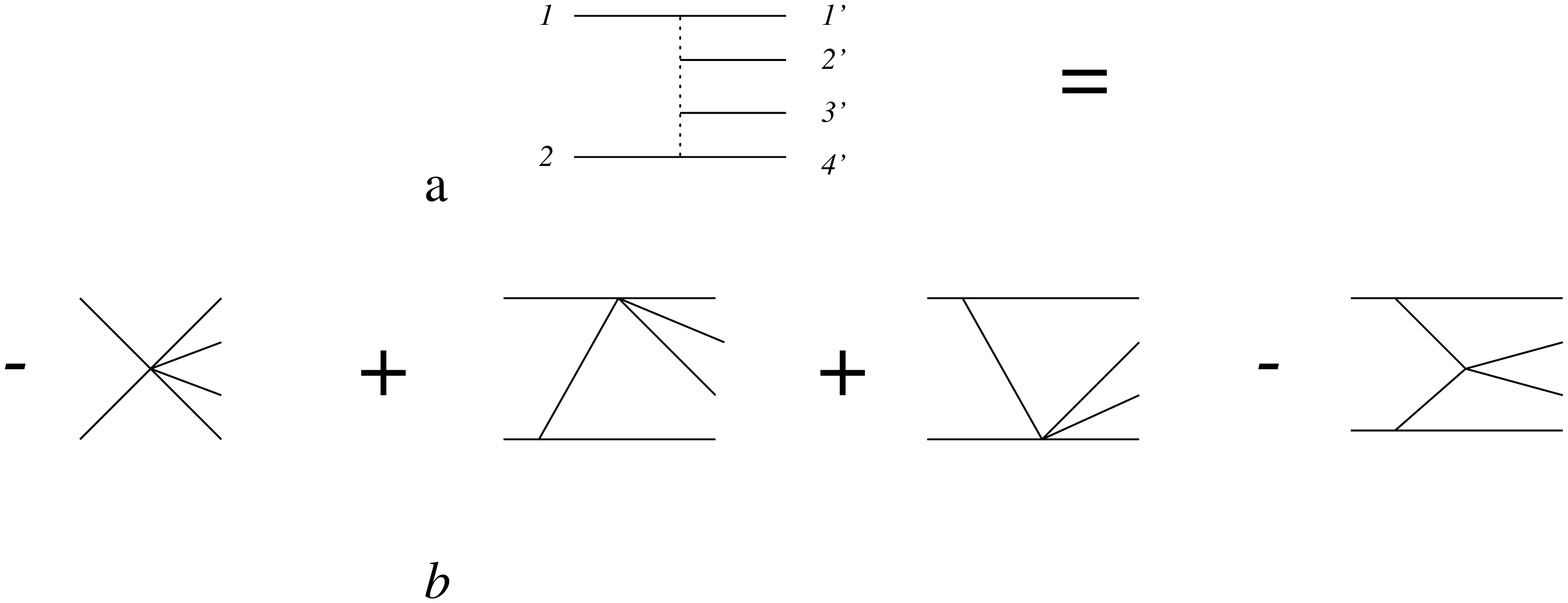, width = 150mm }
\caption{\em
The transition kernel $2 \rightarrow 4$
in s-channel (a) and t-channel (b) representations. }
\end{figure}

\vspace{3cm}

\begin{figure}[thb]
\centering
\epsfig{ file = 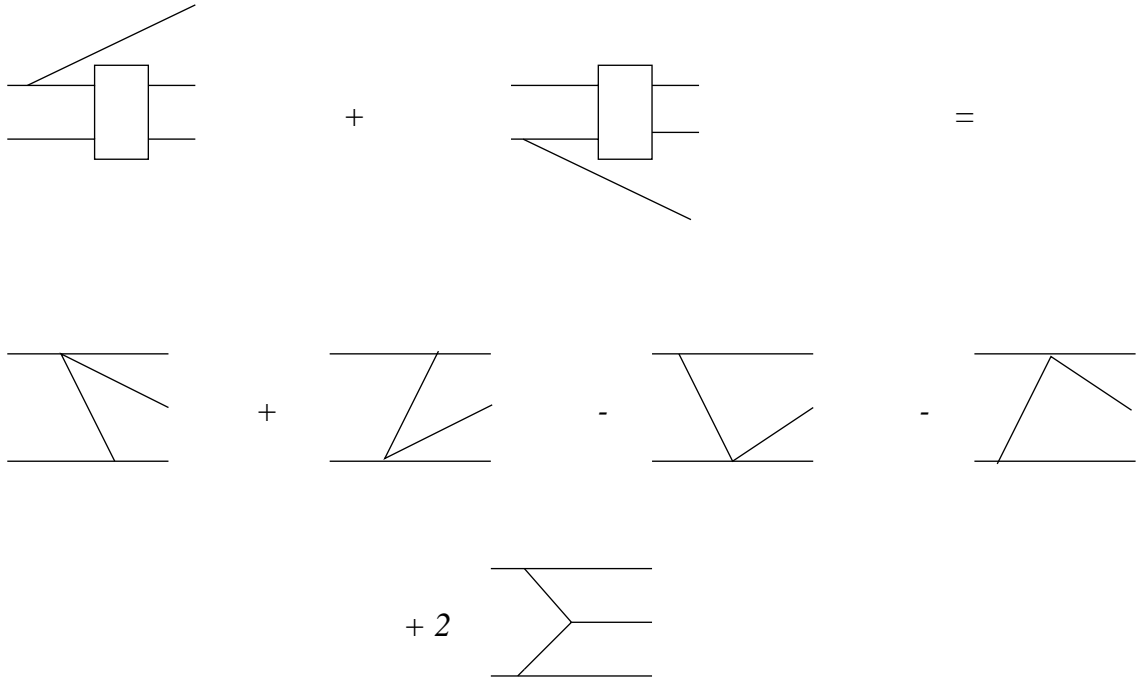, width = 150mm }
\caption{\em
 Contraction of the $2 \rightarrow 3$ splitting with the
          $K_{2 \rightarrow 2}$ interaction.  }
\end{figure}

\vspace{3cm}

\begin{figure}[thb]
\centering
\epsfig{ file = 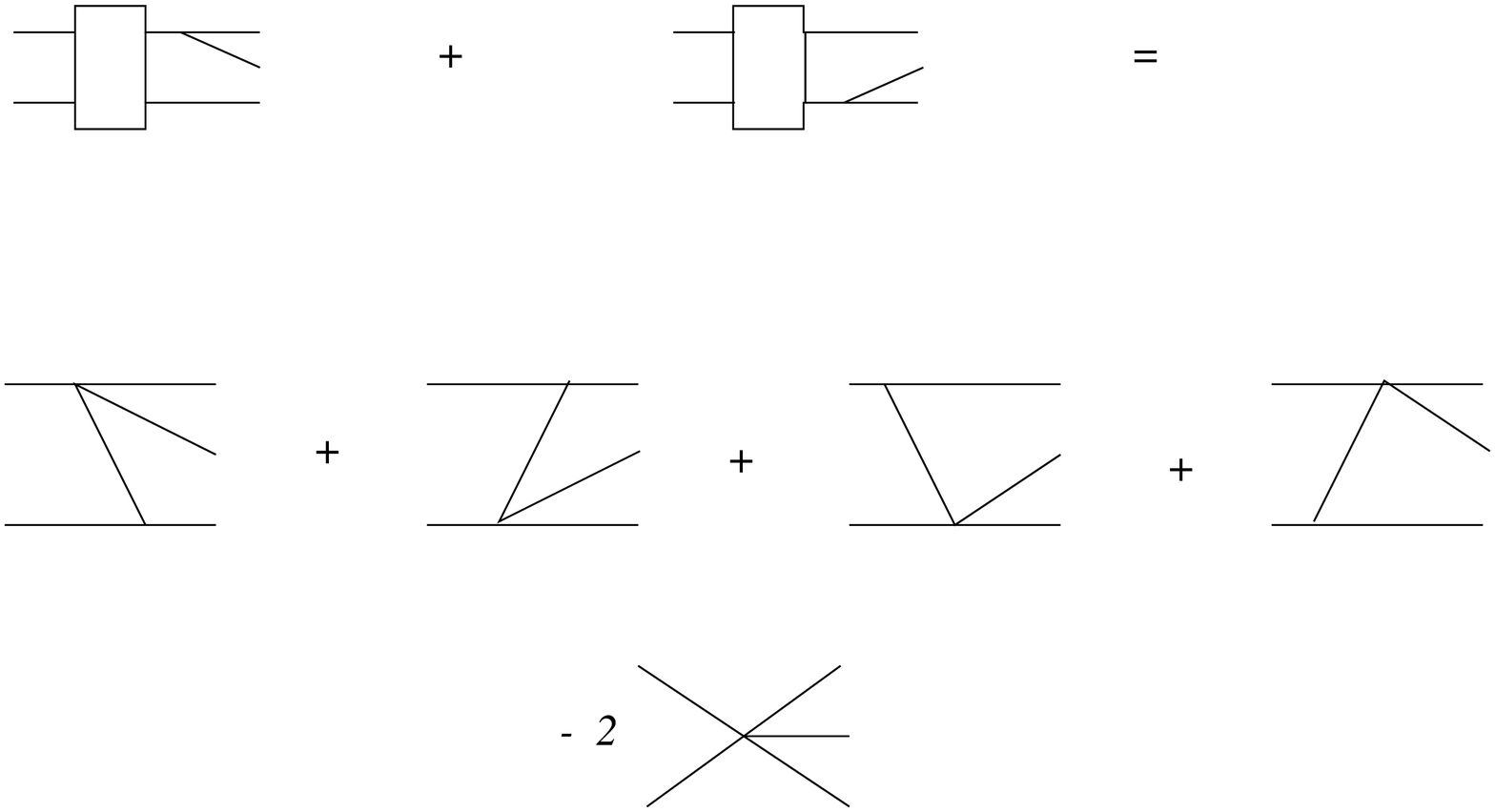, width = 150mm }
\caption{\em
  Contraction of $K_{2 \rightarrow 2}$ with the
          $2 \rightarrow 3$ splitting.  }
\end{figure}

\vspace{3cm}

\begin{figure}[thb]
\centering
\epsfig{ file = 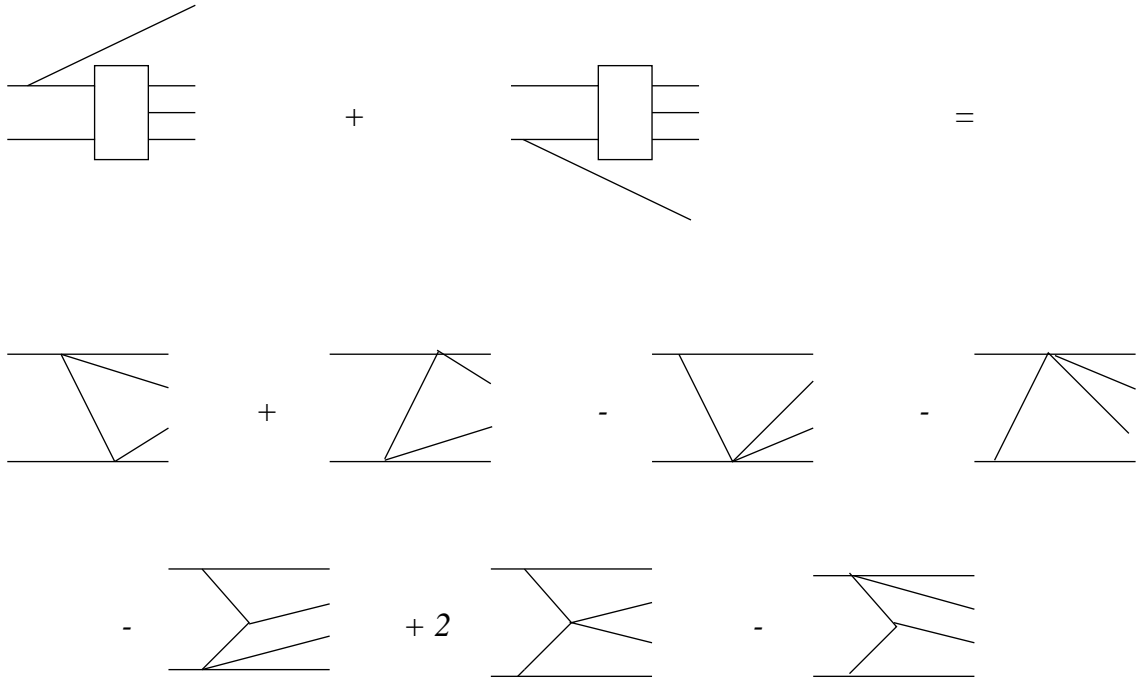, width = 150mm }
\caption{\em
  Contraction of the $2 \rightarrow 3$ splitting with
          the transition kernel $K_{2 \rightarrow 3} $. }
\end{figure}

\vspace{3cm}

\begin{figure}[thb]
\centering
\epsfig{ file = 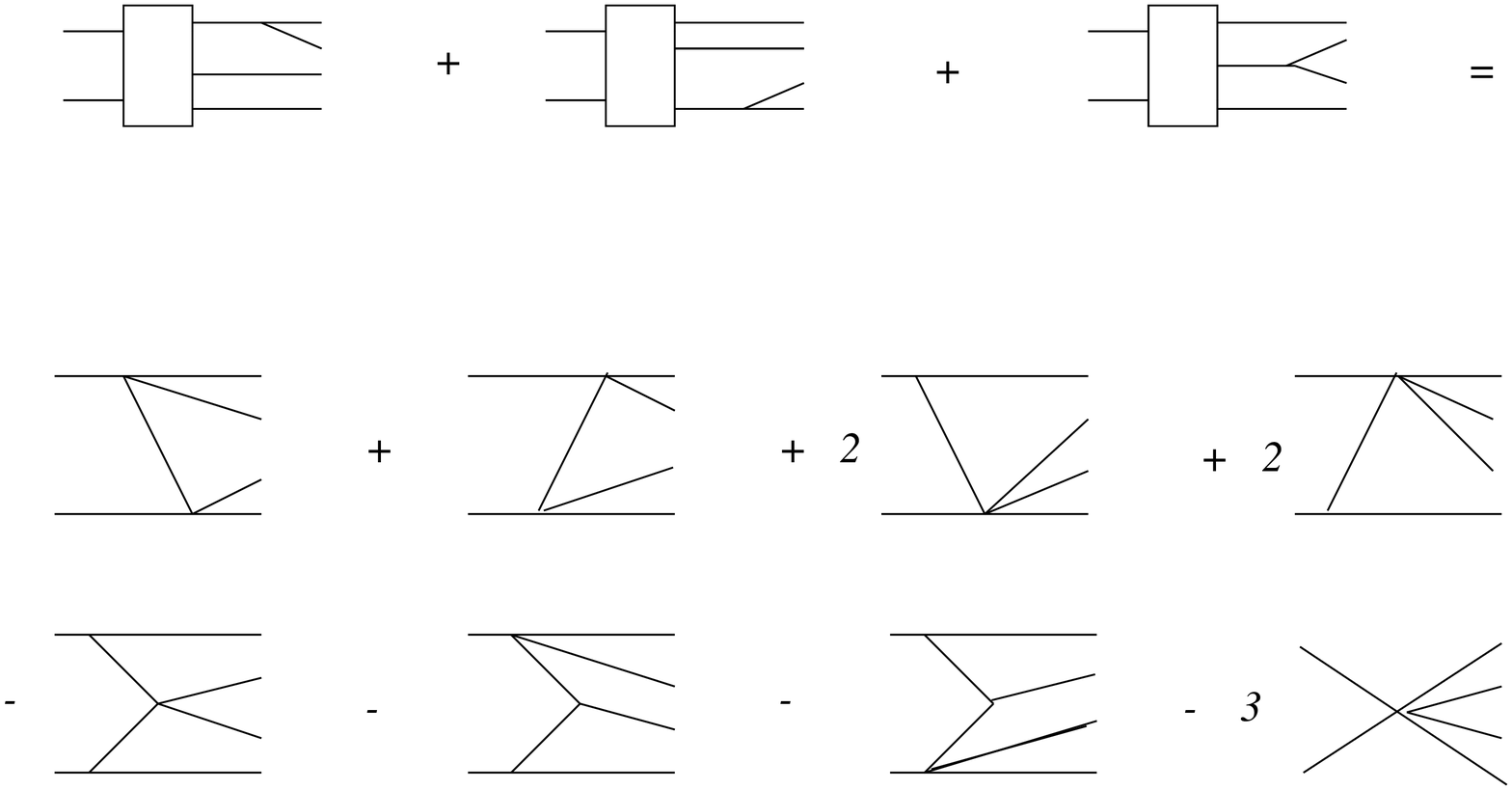, width = 150mm }
\caption{\em
 Contraction of
          the transition kernel $K_{2 \rightarrow 3} $ with the
           $3 \rightarrow 4$ splitting. }
\end{figure}

\vspace{3cm}

\begin{figure}[thb]
\centering
\epsfig{ file = 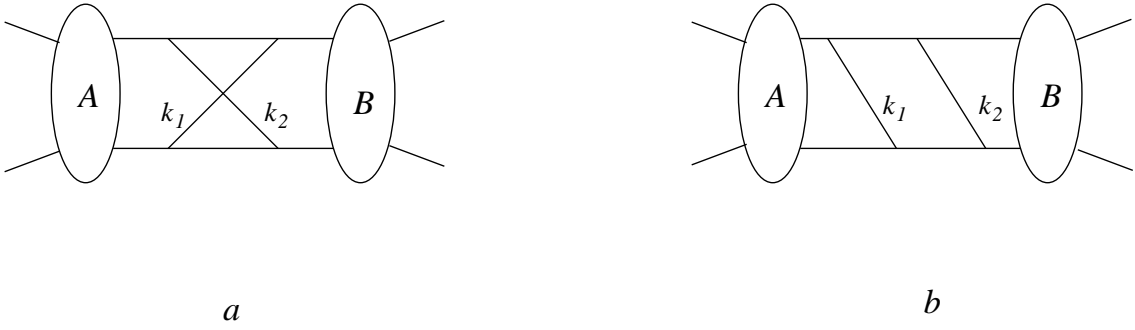, width = 150mm }
\caption{\em
  Contributions to the scattering amplitude. }
\end{figure}
\vspace{3cm}

\begin{figure}[thb]
\centering
\epsfig{ file = 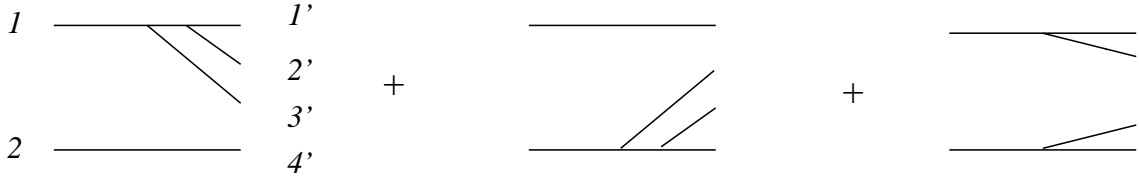, width = 150mm }
\caption{\em
  Graphs contributing to the $2 \rightarrow 4 $ splitting. }
\end{figure}

\vspace{3cm}

\begin{figure}[thb]
\centering
\epsfig{ file = 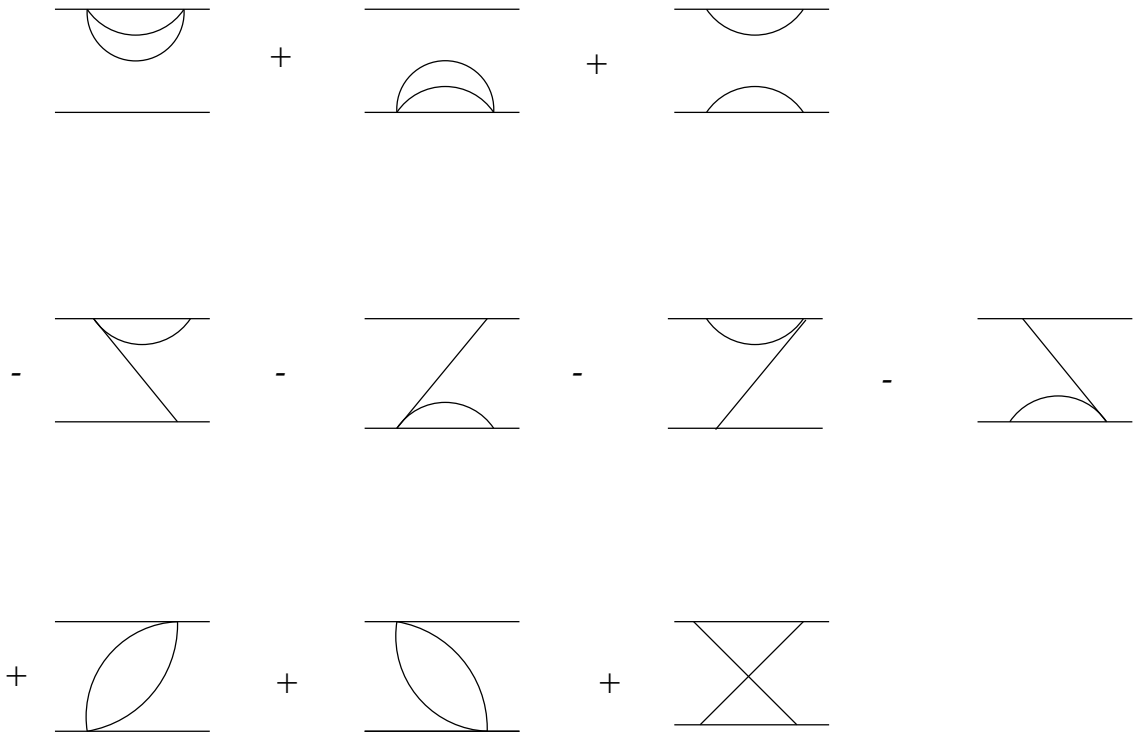, width = 150mm }
\caption{\em
  Transverse momentum graphs for the one-loop correction
        to the $2 \rightarrow 2$ reggeon Green function. }
\end{figure}

\end{document}